\documentclass[aps,twocolumn,showpacs,superscriptaddress,pra,10pt]{revtex4-1}
\usepackage{graphicx}
\usepackage{dcolumn}
\usepackage{bm}
\usepackage{color}
\usepackage{amsmath}
\usepackage{amsfonts}
\usepackage{epstopdf}
\usepackage{subfigure}
\usepackage[makeroom]{cancel}
\usepackage{hyperref}

\begin{document}

\author{Ettore Vitali}
\affiliation{Department of Physics, California State University Fresno, Fresno, California 93740}
\affiliation{Department of Physics, The College of William and Mary, Williamsburg, Virginia 23187}

\author{Patrick Kelly}
\affiliation{Department of Physics, California State University Fresno, Fresno, California 93740}

\author{Annette Lopez}
\affiliation{Department of Physics, California State University Fresno, Fresno, California 93740}
\affiliation{Department of Physics, Brown University, 
Providence, RI 02912}

\author{Gianluca Bertaina}
\affiliation{Dipartimento di Fisica, Universit\`a degli Studi di Milano, via Celoria 16, 20133 Milano, Italy}
\affiliation{Istituto Nazionale di Ricerca Metrologica, Strada delle Cacce 91, 10135 Torino, Italy }

\author{Davide Emilio Galli}
\affiliation{Dipartimento di Fisica, Universit\`a degli Studi di Milano, via Celoria 16, 20133 Milano, Italy}

\title{Dynamical structure factor of a fermionic supersolid on an optical lattice}

\begin{abstract}
Interfacing unbiased quantum Monte Carlo simulations with state-of-art analytic continuation techniques, 
we  obtain exact numerical results for dynamical density and spin correlations in the attractive Hubbard model, describing 
a spin-balanced two-dimensional cold Fermi gas on an optical lattice.
We focus on half-filling, where on average one fermion occupies each lattice site, and the system displays an intriguing supersolid phase: a superfluid with a checkerboard density modulation. The coexistence of $U(1)$ broken symmetry and the density modulations makes this regime very challenging and interesting for the calculation of dynamical properties. We compare our unbiased results with state-of-the-art Generalized Random Phase Approximation calculations: both approaches agree on a well-defined low-energy Nambu-Goldstone collective mode in the density correlations, while the higher energy structures appear to differ significantly. We also observe an interesting high-energy spin mode. We argue that our results provide a robust benchmark for Generalized Random Phase Approximation techniques, which are widely considered to be the method of choice for dynamical correlations in Fermi gases. Also, our calculations yield new physical insight in the high-energy behavior of the dynamical structure factor of the attractive Hubbard model, which is a well known prototype lattice model for superconductors and is a fertile field to target the observation of collective modes in strongly correlated systems. 
\end{abstract}

\maketitle

\section{Introduction}

The study of dynamical response functions of strongly correlated quantum systems is a major challenge in the realm of condensed matter physics, both theoretically and experimentally.
In fact, these quantities are sensitive to the manifold of excited states of a physical system and they allow us to investigate crucial physical observables, like the excitation spectrum and the spectral functions, that cannot be explored in the context of a study of static equilibrium properties. 
In superfluid fermionic systems, the calculation of dynamical correlation functions is particularly exciting. In fact, while for non-interacting fermions the excited states are just the result of particle-hole excitations, the broken $U(1)$-symmetry in Fermi superfluids \cite{Varma2002} is expected to result in two collective modes in the low-energy dynamics of the system: massless fluctuations of the phase of the order parameter (the Nambu-Goldstone mode) and - more elusive - massive fluctuations in the amplitude of the order parameter (the Higgs mode). The study of these modes and of their relation with the particle-hole continuum requires the calculation, or the experimental measurement, of dynamical response functions.
The role of the excited states, which makes dynamical responses such a fertile source of crucial information, makes a fully ab-initio theoretical calculation a formidable challenge. Recently, unbiased calculations of dynamical correlation functions for a few strongly correlated systems have become realistic thanks to the unprecedented progress in computational techniques. In particular, it is now possible to use unbiased quantum Monte Carlo calculations to compute dynamical Green functions and dynamical density correlations \cite{PhysRevB.94.085140,PhysRevA.96.061601} of unpolarized attractive Fermi gases. These results may provide new insight in the physical behavior of correlated systems, both in cold atomic Fermi systems and in lattice models for superconductivity, and they allow us to systematically assess the validity of the state-of-the art theoretical approximation schemes. 

In this paper we address the unbiased calculation of two-body dynamical correlations of the attractive Hubbard model on a square lattice at half-filling using quantum Monte Carlo (QMC) techniques. Our main focus is the density dynamical structure factor $S({\bf{q}},\omega)$, yielding the spectrum of density fluctuations, but we complement the analysis with a study of the spin dynamical structure factor $S^s({\bf{q}},\omega)$, describing spin dynamics in the system. We compare our results with the predictions of generalized Random Phase Approximation (GRPA), which is widely considered a method of choice for the study of dynamical response functions of superfluids and superconductors. GRPA was recently used to investigate strongly correlated Fermi superfluids \cite{PhysRevA.80.043612,PhysRevA.80.063627,Zhao_2020}, including the regime that we are focusing on. Our unbiased results allow us to assess the accuracy of GRPA calculations and to explore effects not accounted for by GRPA. This physical system is a particularly interesting and challenging test ground, as the breaking of the $U(1)$-symmetry giving rise to superfluidity coexists with a checkerboard modulation of the particle density. The Hubbard model in this regime is particularly relevant in cold atomic Fermi systems, perhaps the most accurate laboratory in many-body physics \cite{RevModPhys.80.1215,RevModPhys.80.885,Gross995,PhysRevLett.119.265301,SyntFields,10.1093/ptep/pts031,PhysRevA.97.013601,Brown379,doi:10.1080/00018732.2019.1594094,Mitra_Quantumgasmicroscopy_2018}. In this context, the model describes an intriguing fermionic supersolid. In fact, it has been known for quite a long time that a supersolid phase can be found in fermionic systems. If we confine a fermionic cold gas, made of Lithium or Potassium atoms, in a quasi two-dimensional geometry where an optical lattice is generated using laser standing waves, it is expected that a supersolid state of matter can be realized if the particle density is accurately tuned in such a way that, on average, we have one fermion in each valley of the optical lattice \cite{PhysRevLett.62.1407,PhysRevLett.119.265301}. This condition is referred to as {\it{half-filling}}. Although in the Hubbard model translational symmetry is already broken by the lattice, the term ``supersolid'' is appropriate here since the checkerboard density modulation is a further symmetry breaking.
This system can be thought as the fermionic counterpart of the somewhat more famous bosonic case in the continuum. The latter recently gained a lot of renewed attention. In fact, a supersolid phase was observed in Bose-Einstein condensates coupled with light \cite{Leonard1415,Leonard2017,Li2017} and very exciting theoretical and experimental results were recently published clearly indicating the possibility of realizing a supersolid by engineering a combination of a short-range repulsion and a long-range dipolar interatomic potential \cite{PhysRevLett.122.130405,PhysRevX.9.011051,PhysRevX.9.021012,Natale_ExcitationSpectrumTrapped_2019,Guo_lowenergyGoldstonemode_2019,Tanzi_Supersolidsymmetrybreaking_2019}.

We perform the study by using cutting edge sign-problem free Auxiliary-Field quantum Monte Carlo (QMC) techniques \cite{AFQMC-lecture-notes-2013} interfaced with state-of-the-art analytic continuation techniques \cite{giftREV,gift}. 
We compare the QMC results with GRPA calculations of $S({\bf{q}},\omega)$ and we discuss similarities and differences in order to address the role of many-body correlations not accounted for by the GRPA approximation. We also discuss limitations and advantages of both approaches. 
In our presentation, we also rely on the non-trivial dynamical structure factor of non-interacting fermions on a optical lattice at half-filling, which helps us shed light into the particle-hole continuum of the system, and on the result of a simple BCS theory. 
QMC and GRPA calculations agree very well in detecting a massless Nambu-Goldstone collective mode $\omega({\bf{q}})$. The mode has an evident ``rotonic'' \cite{PhysRevB.46.11025} minimum at ${\bf{q}} = (\frac{\pi}{a},\frac{\pi}{a})$, $a$ being the lattice parameter, whose frequency vanishes in the thermodynamic limit, arising from the crystalline checkerboard order. At higher energy, QMC and GRPA predictions differ significantly. GRPA predictions clearly display a quasi-particle pair continuum at energies higher than $2\Delta$, $\Delta$ being the mean-field superfluid gap of the system.
The QMC results at higher energy, despite some instrumental limitations that will be described below, still allow us to see physical features: there is certainly some renormalization with respect to the GRPA results due to the many-body correlations and the higher energy contribution to $S({\bf{q}},\omega)$ appears to be more narrow. It is quite natural to argue that the QMC is detecting a renormalized quasi-particle pair continuum. On the other hand, our data cannot exclude that a second coherent mode may be present in the dynamical structure factor. Our resolution does not allow us to make strong statements but we cannot exclude the possibility of detecting a Higgs mode, that certainly \cite{Varma2002,PhysRevLett.115.157002,PhysRevB.84.174522,doi:10.1146/annurev-conmatphys-031119-050813,doi:10.1146/annurev-conmatphys-031214-014350} exists but that is expected to decay in particle-holes pairs, or equivalently in pairs of quasi-particles. Also, a second peak has been found in several studies of supersolids \cite{PhysRevLett.108.175301,saccani_bose_2011,PhysRevX.9.021012,Natale_ExcitationSpectrumTrapped_2019,Guo_lowenergyGoldstonemode_2019,Tanzi_Supersolidsymmetrybreaking_2019} and some similar feature may be present in our data.
More resolution will be needed to shed light in this exciting high energy behavior.
We complement the analysis with the study of the spin density structure factor $S^s({\bf{q}},\omega)$, describing the spin dynamics of the system and we find an interesting spin mode close to the upper limit of the quasi-particle pair continuum. The mode is pretty narrow at low momentum, and we discuss possible important implications of its observation.

The paper is structured as follows: in Section II we will introduce the microscopic model of the system and we will briefly describe the QMC method, together with the analytic continuation technique.
In Section III we will present our results for $S({\bf{q}},\omega)$ and we compare our results with GRPA calculations. We also show other important correlation functions for the physical system.
Finally, we will draw our conclusions in Section IV.

\section{Model and method}
We rely on the well known two-dimensional Hubbard model, describing a collection of structureless fermions with spin $1/2$ moving on a lattice. In the realm of cold atoms, the lattice is experimentally realized with standing waves of laser light. The hamiltonian is
\begin{equation}
\label{SecII-HubbardHamiltonian}
\begin{split}
& \hat{H} = \hat{K} + \hat{V} \\
& \hat{K} = -t \sum_{\langle {\bf{r}}, {\bf{r}^{\prime}} \rangle, \sigma}
 \hat{c}^{\dagger}_{{\bf{r}},\sigma} \hat{c}^{}_{{\bf{r}^{\prime}},\sigma} \\
 & \hat{V} =
 U \sum_{{\bf{r}}} \hat{n}_{\uparrow}({\bf{r}})\hat{n}_{\downarrow}({\bf{r}})
 \end{split}
\end{equation}
with:
\begin{equation}
\hat{n}_{\sigma}({\bf{r}}) = \hat{c}^{\dagger}_{{\bf{r}},\sigma} \hat{c}^{}_{{\bf{r}}, \sigma} \,.
\end{equation}
In Eq.~\eqref{SecII-HubbardHamiltonian}, $\bf{r}$ runs over the sites of a square lattice made of $\mathcal{N}_s = L \times L$ sites, defined by the minima of the optical lattice, while $\sigma$ is the spin orientation. The first term in the Hamiltonian Eq.~\eqref{SecII-HubbardHamiltonian} is the kinetic energy, describing particles hopping among nearest-neighbor lattice sites with amplitude $t$, while the second term is the on-site interaction, whose strength is given by the parameter $U$. In this paper, we will focus on the attractive case, and thus $U < 0$.
The symbol $\langle {\bf{r}}, {\bf{r}^{\prime}} \rangle$ means that the sites are nearest neighbors. We will use periodic boundary conditions throughout this paper. The Hubbard model \cite{Hubbard_1,Hubbard_2} is one of the most widely studied models in atomic physics and in condensed matter physics. Despite its apparent simplicity, no analytical solutions to this hamiltonian are known in more than one dimension.
Although in most applications the Hubbard model has a repulsive interaction, related to Coulomb force, also the attractive case is extremely interesting, both in the realm of cold atoms and in condensed matter physics \cite{RevModPhys.62.113}.

The main purpose of this paper is to compute the dynamical structure factor of the system at zero temperature:
\begin{equation}
\label{SecII-Sqo}
S({\bf{q}},\omega) = \frac{1}{ N} \int_{-\infty}^{+\infty} \frac{dt}{2\pi} \, e^{i\omega t} \,
 \left\langle \Psi_0 \, | e^{i t \hat{H}} \hat{n}_{{\bf{q}}} e^{- i t \hat{H}}\, \hat{n}_{{-\bf{q}}} | \Psi_0 \right\rangle 
\end{equation}
where $| \Psi_0 \rangle$ is the ground state of \eqref{SecII-HubbardHamiltonian}, while $\hat{n}_{{\bf{q}}} $ is the Fourier component of the density of particles:
\begin{equation}
\hat{n}_{{\bf{q}}} = \sum_{{\bf{k}}, \sigma} \hat{c}^{\dagger}_{{\bf{k}},\sigma} \hat{c}^{}_{{\bf{k}}+{\bf{q}},\sigma} \,.
\end{equation}
All the momenta, ${\bf{q}}$ and ${\bf{k}}$ belong to the first Brillouin zone of the square lattice, $[-\frac{\pi}{a}, \frac{\pi}{a}) \times [-\frac{\pi}{a}, \frac{\pi}{a})$. All the lengths in this paper will be measured in units of $a$ and, therefore, we will set $a=1$ from now on. 
Moreover, in the discussion of the dynamical properties, we will implicitly assume $\hbar=1$, as usual in the studies of the Hubbard model. 
Since our numerical approach requires us to work with finite lattices of linear size $L$ and we choose periodic boundary conditions, the momenta are discretized: ${\bf{k}} = \frac{2\pi}{L} {\bf{n}}$, where ${\bf{n}} \in \mathbb{Z}^2$.

Computing Eq.~\eqref{SecII-Sqo} for a correlated quantum system is a huge challenge: in this work, we will use the cutting-edge Auxiliary-Field QMC technique to sample at the same time both the ground state wave function of the system $| \Psi_0 \rangle$ and the propagator in imaginary time $\exp(- \tau \hat{H} )$: this will allow us to compute exactly the intermediate scattering function in imaginary time, as discussed below. We will then use the genetic inversion via falsification of theories (GIFT) to perform the analytic continuation necessary to compute $S({\bf{q}},\omega)$.

The Auxiliary-Field QMC belongs to a class of QMC techniques that have been known for a long time to be methods of choice for the study of the attractive Hubbard model for superconductors (see for example \cite{PhysRevB.54.1286}).
In particular, we exploit recent methodological advancements in the Auxiliary-Field QMC technique that allow us to compute exact dynamical correlations in imaginary time with a very favorable scaling as a function of the size of the system \cite{PhysRevA.92.033603,PhysRevA.96.061601,PhysRevB.94.085140}. The analytic continuation, which is necessary to estimate properties in real time, in particular $S({\bf{q}},\omega)$, is performed using the Genetic Inversion via Falsification of Theories (GIFT) technique \cite{giftREV,gift}, which has been proved to provide robust results for quantum many-body systems \cite{PhysRevA.96.061601,helium1D,rods1D,arrigoni_excitation_2013,he3_dyn}.

The Auxiliary-Field QMC technique relies on the projection formula:
 \begin{equation}
 \label{SectII-proj}
| \, \Psi_0 \rangle \propto \lim_{\beta \to +\infty} e^{-\beta ( \hat{H} - E_0)} |\phi_0\rangle
\end{equation}
which allows us to asymptotically project a given approximation $|\phi_0\rangle$ to the ground state wave function of the system onto the ground state itself.
In \eqref{SectII-proj} $E_0$ is an estimate of the ground state energy while the approximation $|\phi_0\rangle$ is chosen to be a Slater determinant with $N_{\uparrow}$ spin-up and $N_{\downarrow}$ spin-down particles, such that 
 $|\phi_0\rangle$ is not orthogonal to the $\mathcal{N}_p$-particle ($\mathcal{N}_p=N_{\uparrow}+N_{\downarrow}$) ground state $ | \,\Psi_0 \, \rangle$ of \eqref{SecII-HubbardHamiltonian}. 
 In the simplest case, $|\phi_0\rangle$ is simply the ground state wave function of the non-interacting Hubbard model. In this work we will focus on the behavior of the model at half-filling, which means that $\mathcal{N}_p = L \times L$, and without spin polarization, meaning $N_{\downarrow}=N_{\uparrow}$.
 
Starting from a Trotter decomposition:
\begin{equation}
e^{-\beta ( \hat{H} - E_0)} = \left(e^{-\delta\tau ( \hat{H} - E_0)} \right)^{M}, \quad \delta\tau = \frac{\beta}{M}
\end{equation}
which allows us to rely on the behavior of the propagator for small imaginary time $\delta\tau$, the Auxiliary-Field QMC method uses an Hubbard-Stratonovich transformation which yields the following expression:
 \begin{equation}
 \label{SecII-propagator}
e^{-\delta\tau ( \hat{H} - E_0)} \simeq \int d{\bf{x}} \, p({\bf{x}}) \, \hat{B}({\bf{x}}) \quad \delta\tau \to 0 \quad.
\end{equation}
In Eq.~\eqref{SecII-propagator} the integrations runs over all the possible configurations of an auxiliary field ${\bf{x}}$ defined on the lattice: in our case, ${\bf{x}}$ is an Ising field, $x({\bf{r}})=\pm 1$.
The function $p({\bf{x}})$ is a uniform probability density
on the space of the configurations of the auxiliary field: $p({\bf{x}})= \frac{1}{2^{\mathcal{N}_p}}$.
An explicit expression for the operator $\hat{B}(\bf{x})$ appearing in \eqref{SecII-propagator} is given by the following \cite{PhysRevB.28.4059}:
\begin{equation}
\label{SecII-operatorB}
\hat{B}({\bf{x}}) = e^{\delta\tau E_0} \, e^{-\delta\tau\hat{K}/2} \, \prod_{{\bf{r}}} \hat{b}_{{\bf{r}}}(x({\bf{r}}))\,e^{-\delta\tau\hat{K}/2} 
\end{equation}
where $\hat{K}$ is the kinetic energy of the Hubbard model, while:
\begin{equation}
 \hat{b}_{{\bf{r}}}(x) = e^{-\delta\tau U (\hat{n}({\bf{r}}) - 1)/2 } \, e^{\gamma x (\hat{n}({\bf{r}}) - 1)}
\end{equation}
with $\hat{n}({\bf{r}}) = \hat{n}_{\uparrow}({\bf{r}}) + \hat{n}_{\downarrow}({\bf{r}})$ being the particle density operator while $\cosh(\gamma)=\exp\left(\frac{\delta\tau |U|}{2}\right)$. From the point of view of the physical interpretation, the Hubbard-Stratonovich transformation allows us to map the interacting problem onto an ensemble of non-interacting systems of fermions moving in random external potentials.
The average over the ensemble recovers the fully interacting model.
From the point of view of implementation, the key point of the methodology is the fact that the operator $\hat{B}(\bf{x})$
defined in Eq.~\eqref{SecII-operatorB} is the exponential of a one-body operator dependent on the auxiliary field configuration. This implies that the operator $\hat{B}(\bf{x})$ maps Slater determinants into Slater determinants.
This allows the QMC procedure to implement a random walk whose state space coincides with the manifold of Slater determinants with $\mathcal{N}_p$ particles.
Such a random walk opens the possibility to compute expectation values of any physical property, say $\hat{O}$, using Monte Carlo integration, relying on the expression:
\begin{equation}
\langle \, \hat{O} \, \rangle 
\simeq \frac{\int d{\bf{x}}_1 \dots d{\bf{x}}_{2M} \, \pi({\bf{x}}_0, \dots, {\bf{x}}_{2M}) \mathcal{O}({\bf{x}}_0, \dots, {\bf{x}}_{2M})}
{\int d{\bf{x}}_1 \dots d{\bf{x}}_{2M} \, \pi({\bf{x}}_0, \dots, {\bf{x}}_{2M})}
\end{equation}
where $\langle \, \hat{O} \, \rangle$ is shorthand for $\langle \Psi_0 \, | \, \hat{O} \, | \, \Psi_0 \rangle$, while:
\begin{equation}
\begin{split}
& \pi({\bf{x}}_1, \dots, {\bf{x}}_{2M}) = \prod_{i=1}^{2M} p({\bf{x}}_i) \langle \phi_{L} \, | \phi_{R} \rangle \\
& \mathcal{O}({\bf{x}}_1, \dots, {\bf{x}}_{2M}) = \frac{\langle \phi_{L} \, | \hat{O} \, | \, \phi_{R} \rangle}{\langle \phi_{L} \, | \phi_{R} \rangle} \,.
\end{split}
\end{equation}
The ``left'' and ``right'' Slater determinants are defined as:
\begin{equation}
\begin{split}
& \langle \phi_{L} \, | = \langle \phi_{0} \, | \, \hat{B}({\bf{x}}_{2M}) \dots \hat{B}({\bf{x}}_{M+1}) \\
& | \phi_{R} \rangle =\, \hat{B}({\bf{x}}_{M}) \dots \hat{B}({\bf{x}}_{1}) | \phi_{0} \rangle \,.
\end{split}
\end{equation}
In the simplest implementation of the methodology, we use the non-interacting wave function as the initial wave function:
\begin{equation}
| \phi_{0} \rangle = \prod_{{\bf{k}},\sigma \,\, \varepsilon({{\bf{k}}}) < \varepsilon_F} \hat{c}^{\dagger}_{{\bf{k}},\sigma} |0 \rangle
\end{equation}
where $\varepsilon({{\bf{k}}}) = -2t (\cos(k_x) + \cos(k_y) )$ is the dispersion relation of the non-interacting two-dimensional Hubbard model, $\varepsilon_F$ is the Fermi energy, while the operator:
\begin{equation}
\hat{c}^{\dagger}_{{\bf{k}},\sigma} = \frac{1}{\sqrt{\mathcal{N}_s}}\sum_{{\bf{r}}} e^{i {\bf{k}} \cdot {\bf{r}}} \, \hat{c}^{\dagger}_{{\bf{r}},\sigma}
\end{equation}
creates one particle with spin orientation $\sigma$ in a plane wave $e^{i {\bf{k}} \cdot {\bf{r}}}/{\sqrt{\mathcal{N}_s}}$.
In the the state $| \phi_{0} \rangle$, all the orbitals of the particles in the system will be plane-waves.
The Slater determinant $| \phi_{R} \rangle$, for a given imaginary time dependent configuration of the auxiliary-field $({\bf{x}}_1, \dots, {\bf{x}}_{2M})$ will be obtained from $| \phi_{0} \rangle$, by 
applying the product of the operators $\hat{B}({\bf{x}}_{i})$ to the plane waves, once for each particle in the system.
Similarly we build $ \langle \phi_{L} \, |$. Each Slater determinant $| \phi \rangle $ is parametrized as $\Phi=\Phi^{\uparrow} \otimes \Phi^{\downarrow} $ where $\Phi^{\sigma}$ is a $\mathcal{N}_{s} \times N_{\sigma}$ complex matrix, containing all the components $\langle {\bf{r}}, \sigma | \phi \rangle$. The matrix elements $\mathcal{O}$ are obtained with simple linear algebra manipulations. This is the essence of the QMC method: we randomly sample imaginary time dependent configurations of the auxiliary-field and this allows us to implement a random walk in the manifold of the Slater determinants for $\mathcal{N}_p$ particles; an average over all the possible random walks yields physical properties of the correlated systems. 
Several technical improvements can make the technique more efficient, including importance sampling and force bias \cite{PhysRevA.92.033603,PhysRevA.100.023621}. The resulting methodology has a very favorable scaling as a function of the size of the system: precisely it scales as $\mathcal{O}(\mathcal{N}_p^2 \mathcal{N}_s ).$
It important to mention that, for the spin-balanced ($N_{\uparrow} = N_{\downarrow}$) attractive Hubbard model, the method is sign-problem free, which implies that we can compute exact physical properties of the system. By exact we mean that, for a given choice of $\mathcal{N}_p$ and $\mathcal{N}_s$, we can always choose the parameters of the quantum Monte Carlo run, and namely the time step $\delta \tau$ and the total projection time $\beta = M \delta \tau$, in such a way that the systematic error is smaller than the statistical uncertainty for a given computation time.
The Auxiliary-Field QMC method can be extended to the calculation of dynamical properties. As described in detail in the papers \cite{PhysRevB.94.085140,PhysRevA.96.061601}, it is possible to compute exactly dynamical correlations in imaginary time, like the intermediate scattering function:
\begin{equation}
F({\bf{q}},\tau) = \left\langle \Psi_0 \, | e^{\tau \hat{H}} \hat{n}_{{\bf{q}}} e^{- \tau \hat{H}}\, \hat{n}_{{-\bf{q}}} | \Psi_0 \right\rangle
\end{equation}
without effecting the favorable scaling of the methodology.
This is easily achieved since the imaginary time evolution operator $e^{- \tau \hat{H}}$ is the same operator that allows us to sample the ground state wave function using the projection formula \eqref{SectII-proj}, and the same formal manipulations can be applied to it. This results in an efficient algorithm to compute $F({\bf{q}},\tau)$, with the same complexity that is required by static calculations.
We use the intermediate scattering function from quantum Monte Carlo as an input for the analytic continuation problem, that is needed to estimate the dynamical structure factor $S({\bf{q}},\omega)$ through the relation:
\begin{equation}
\label{Laplace}
F({\bf{q}},\tau) = \int_{0}^{+\infty} d\omega \, e^{-\tau \omega} S({\bf{q}},\omega) \,.
\end{equation}
We stress once more that our calculations of $F({\bf{q}},\tau)$ are unbiased, as no sign problem exists for the spin-balanced attractive Hubbard model in the framework of the Auxiliary-Field QMC methodology.
We use the state-of-art Genetic Inversion via Falsification of Theories (GIFT) \cite{gift,giftREV} method to estimate the dynamical structure factor $S({\bf{q}},\omega)$ starting from $F({\bf{q}},\tau)$. We improve the accuracy of the procedure by including the $f$-sum rule, which will be discussed in appendix~\ref{appA}. 
This method has been widely employed to study dynamical properties of superfluid $^4$He in different dimensionalities \cite{gift,giftREV,helium1D,arrigoni_excitation_2013}, normal $^3$He \cite{he3_dyn}, cold atomic systems \cite{PhysRevA.96.061601,rods1D,PhysRevLett.119.215301,PhysRevLett.108.175301} and the Hubbard model \cite{PhysRevB.94.085140}, and it is known to provide robust results.
In the following section we will present our results.

\begin{figure}[ptb]
\includegraphics[width=\columnwidth, angle=0]{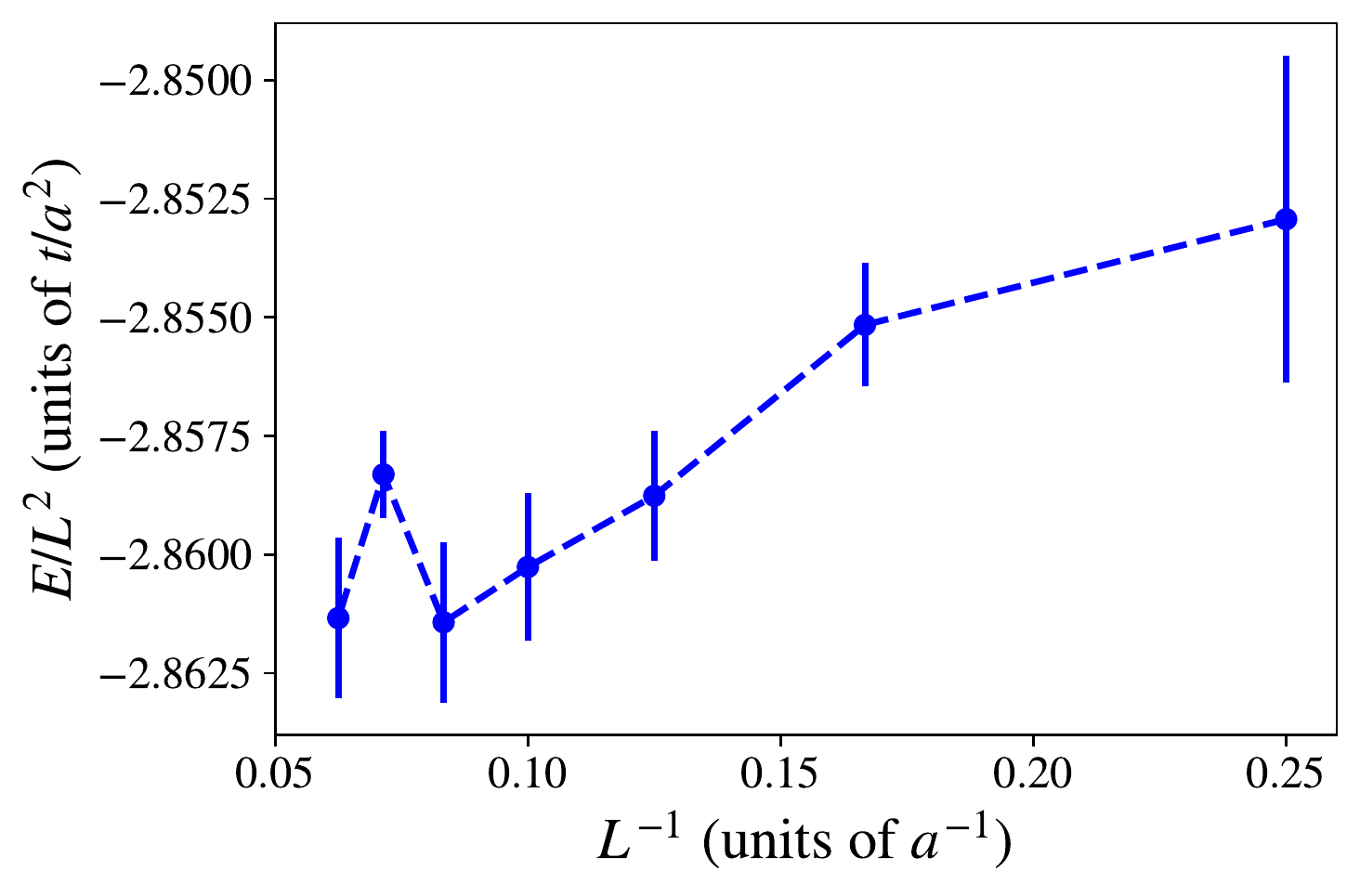}
\caption{(Color online) Energy per site of the Hubbard model at half-filling for $U/t = -4$ as a function of $\frac{1}{L}$, where $L$ is the linear size of the system. The energies are in units of the hopping amplitude $t$.} 
\label{fig:spin-balanced}
\end{figure}

\begin{figure*}[tbp]
\includegraphics[width=\textwidth]{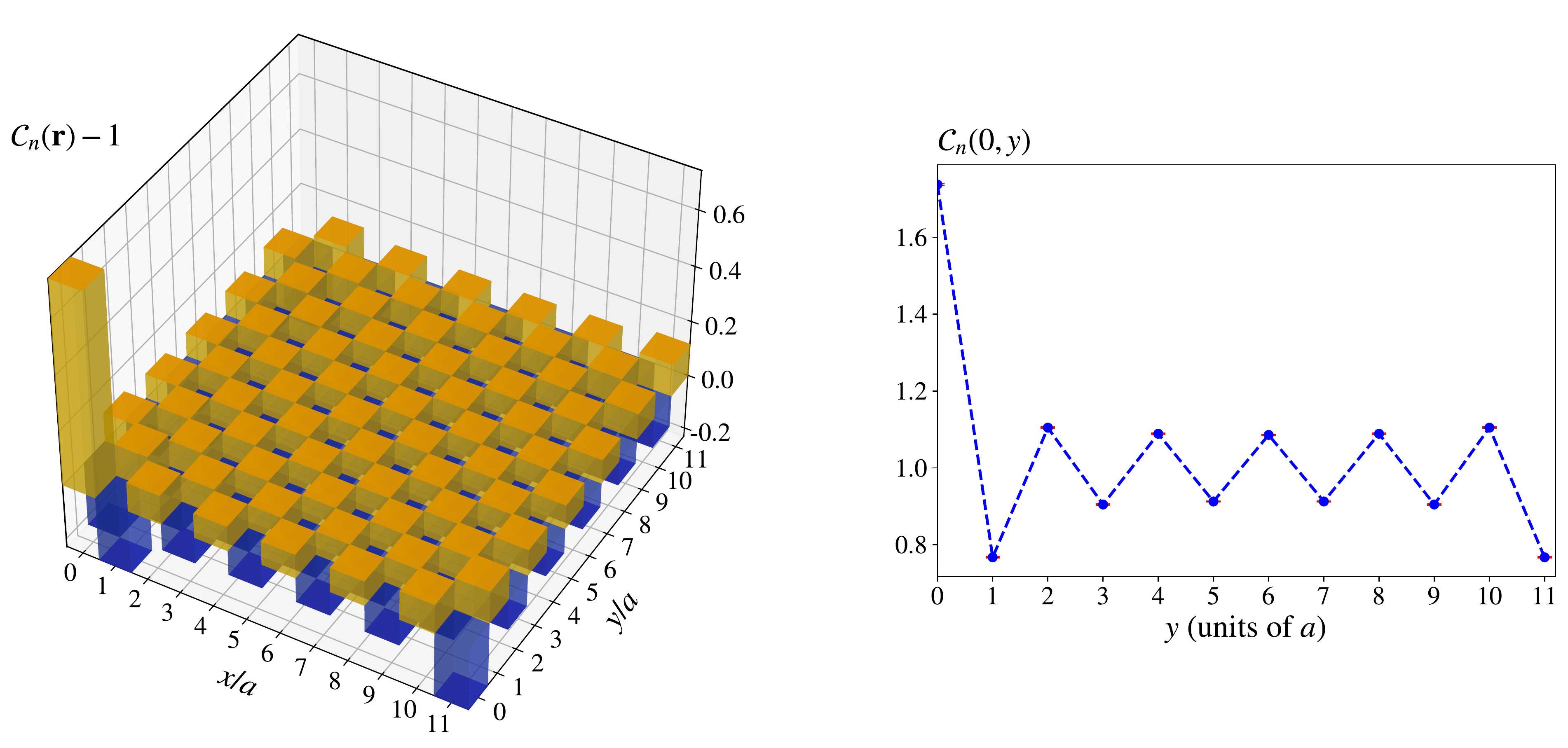}
\caption{(Color online) Left panel: offset of the density correlation $\mathcal{C}_{n}({\bf{r}})$ (dimensionless) with respect to $1$, for the attractive Hubbard model with $U/t=-4$ as a function of the position ${\bf{r}}=(x, y)$. Two colors have been chosen to aid the reader: red (light gray) for lattice sites with density correlation larger than $1$, and blue(dark gray) for lattice sites with density correlation less than $1$. Right panel: a slice of the density correlation taken with $x = 0$. 
The uncertainties are plotted but may be difficult to see, as they are consistently below the size of the symbols. }.
\label{fig:densitycombin} 
\end{figure*}

\section{Results}
We study the attractive Hubbard model on a square lattice with $\mathcal{N}_s=L \times L$ sites hosting $\mathcal{N}_p$ particles at half-filling, that is $\mathcal{N}_p=\mathcal{N}_s$ and spin balance. The strength of the interaction is chosen to be $U/t = -4$. 
Since we target bulk properties, it is crucial to perform size extrapolation, in order to make sure that the size effects are below the level of the statistical uncertainties in the Monte Carlo calculations.
In Fig.~\ref{fig:spin-balanced} we plot the energy per site as a function of the linear size, for $L=4, 6, 8, 10, 12,14, 16$. The results clearly show that, within an uncertainty of the order of $10^{-3}$ the energy per site converges to its bulk limit around $L=10$. All the calculations that we present are for $L=12$. All the other parameters of the simulations, like time-step and total projection time are tuned in such a way that the systematic error is below the level of the uncertainties in the Monte Carlo data.

Before presenting our calculations of the dynamical properties, which are the central product of this work, we present our results about the density correlation and the pairing correlations, defined respectively as:
\begin{equation}
\mathcal{C}_{n}({\bf{r}}) = \left\langle \Psi_0 \, | \, \hat{n}({\bf{r}}) \hat{n}(0) \, | \, \Psi_0 \right\rangle
\end{equation}
where $\hat{n}({\bf{r}}) = \hat{c}^{\dagger}_{{\bf{r}},\uparrow} \hat{c}^{}_{{\bf{r}},\uparrow} + \hat{c}^{\dagger}_{{\bf{r}},\downarrow} \hat{c}^{}_{{\bf{r}},\downarrow} $ and:
\begin{equation}
\mathcal{C}_{pair}({\bf{r}}) = \left\langle \Psi_0 \, | \, \hat{\Delta}^{\dagger}({\bf{r}}) \hat{\Delta}(0) \, | \, \Psi_0 \right\rangle
\end{equation}
where the on-site pairing is defined as:
\begin{equation}
\hat{\Delta}^{}({\bf{r}}) = \hat{c}^{}_{{\bf{r}},\downarrow} \hat{c}^{}_{{\bf{r}},\uparrow} \,.
\end{equation} 
The density correlations, shown in Fig.\ref{fig:densitycombin}, exhibit clear long-range order.
The system is in a crystalline phase, with a checkerboard modulation of the particle of wave vector ${\bf{q}}=(\pi,\pi)$; that is, the order parameter of the solid phase has the form:
\begin{equation}
n({\bf{r}}) = 1 + A \cos({\bf{q}} \cdot {\bf{r}}), \, \, A < 1.
\end{equation}
We observe that this is not the trivial order imposed by the optical lattice: it is a density modulation that arises from the interplay between interatomic correlations and nesting of the non-interacting Fermi surface at half-filling, as we we will discuss in more detail below.

\begin{figure}[btp]
\includegraphics[width=\columnwidth]{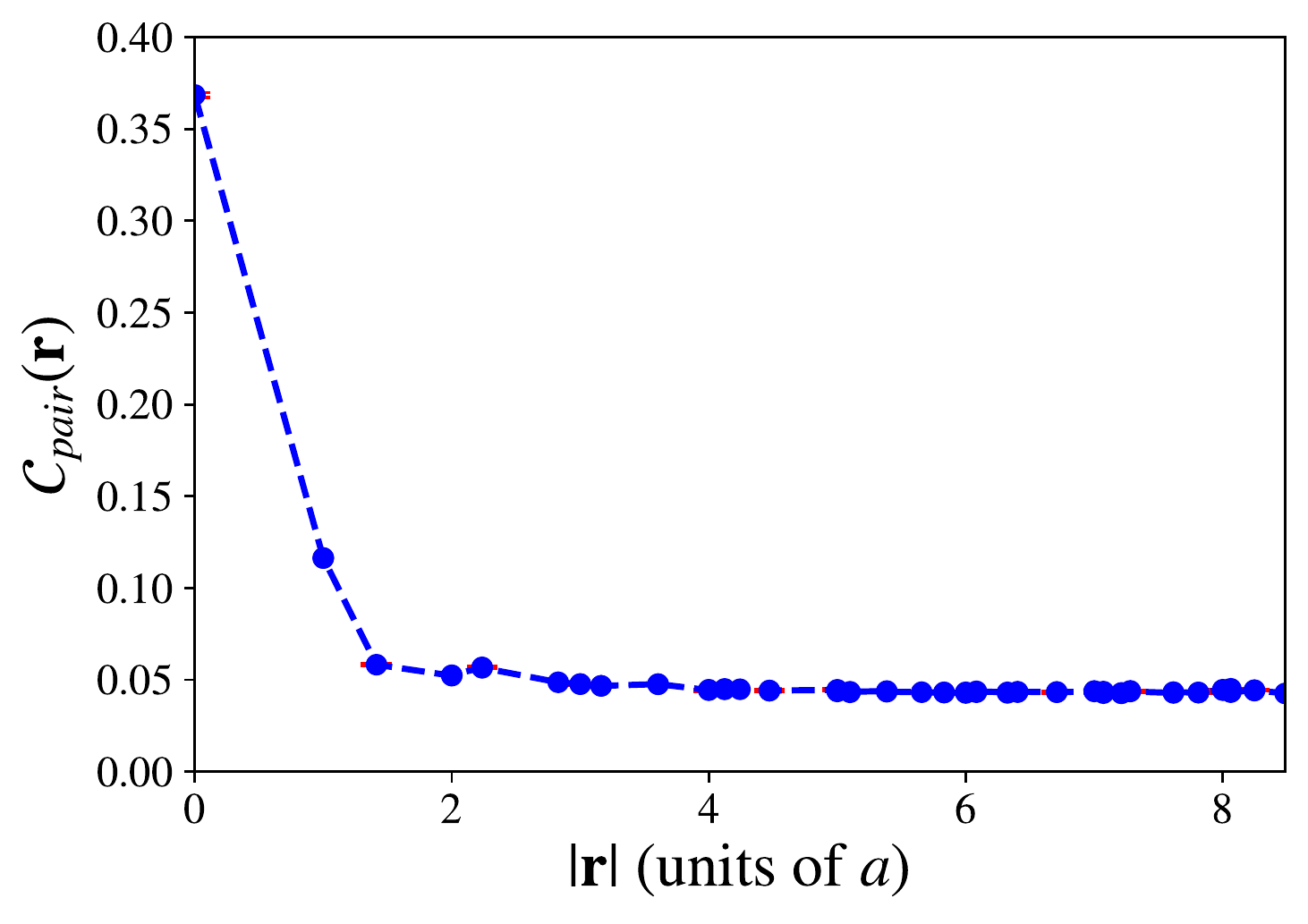}
\caption{(Color online) Pairing Correlation $\mathcal{C}_{pair}({\bf{r}})$ (dimensionless) for the attractive Hubbard model at half-filling with $U/t=-4$ as a function of the distance $|{\bf{r}}|$. The uncertainties are plotted but may be difficult to see, as they are consistently below the size of the symbols.}
\label{fig:pairing}
\end{figure}

The pairing correlation is shown in Fig.\ref{fig:pairing} as a function of distance $|{\bf{r}}|$ and clearly displays convergence to a non-zero limit: 
\begin{equation}
\mathcal{C}_{pair}({\bf{r}}) \to n_0^2 > 0, \quad |{\bf{r}}| \to +\infty.
\end{equation}
This indicates the emergence of off-diagonal long range order, corresponding to a superfluid phase.
The order parameter $n_0$ plays the role of a condensate fraction. Intuitively, the singlet-pairs that form in the system as a consequence of the attractive interaction become coherent and form the equivalent of a Bose-Einstein condensate. We observe that the pairs have total spin equal to zero, and thus obey Bose statistics. 
The combination of the long-range orders in both the density and pairing correlations is the signature of a supersolid phase, which appears to be the equilibrium state at half-filling.
We comment that the coexistence of a condensate fraction and of a nontrivial checkerboard density pattern justifies the name of supersolid for this system.
We stress that our calculations are unbiased, and can serve as useful benchmark for future studies.

\begin{figure*}[tp]
\begin{center}
\includegraphics[width=\textwidth]{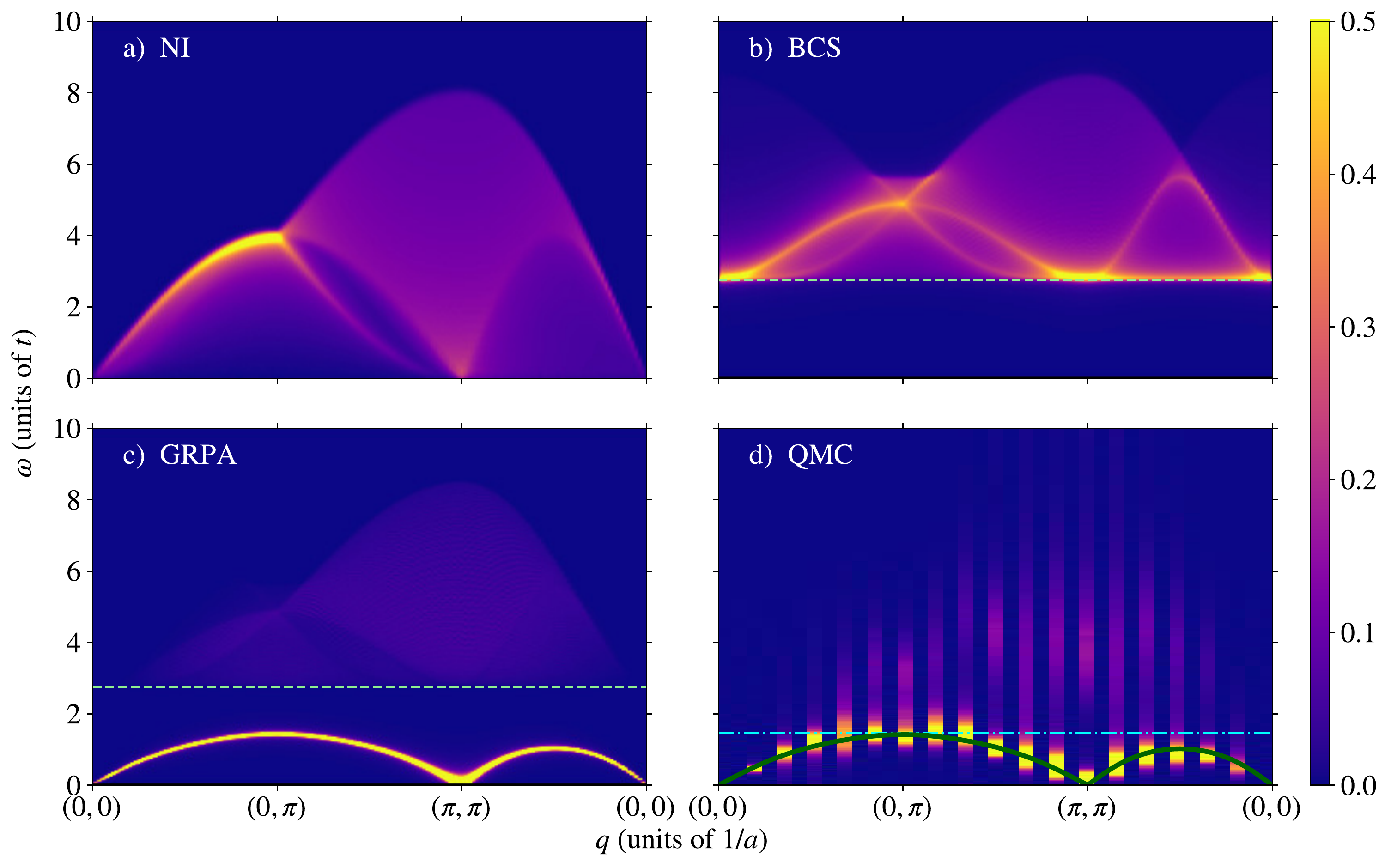}
\caption{(Color online) Panel (a): color plot of the dynamical structure factor $S_0({\bf{q}}, \omega)$ (arbitrary units) for the non-interacting balanced Fermi model at half-filling as a function of momentum ${\bf{q}}$ along a triangle in the first Brillouin zone and frequency $\omega$, in units of the hopping amplitude $t$. Panel (b): color plot of $S({\bf{q}}, \omega)$ (arbitrary units) for the attractive Hubbard model with $U/t=-4$ at the same density as in panel (a) within BCS theory. Dashed horizontal line: twice the quasi-particle gap within BCS theory. Panel (c): color plot of $S({\bf{q}}, \omega)$ (arbitrary units) for the same system within GRPA. Dashed horizontal line: twice the quasi-particle gap within GRPA. Panel (d): color plot of $S({\bf{q}}, \omega)$ (arbitrary units) for the same system calculated with QMC. Solid line: main GRPA mode from panel (c). Dot-dashed horizontal line: twice the quasi-particle gap calculated with QMC from Ref.~\cite{PhysRevB.94.085140}.}
\label{fig:Sqomega}
\end{center}
\end{figure*}

The central result of our paper is the QMC calculation of the dynamical structure factor $S({\bf{q}},\omega)$ of the system in this unique phase and the comparison with the predictions of the state-of-the art GRPA approximation.
Together with the QMC calculation, we also estimated $S^{GRPA}({\bf{q}},\omega)$, by using the method detailed in ~\cite{PhysRevA.80.043612}, which allows us to compute the density response function $\chi_{nn}({\bf{q}},\omega)$. The latter is defined, as usual, by the linear response relation:
\begin{equation}
\delta n({\bf{q}},\omega) = \chi_{nn}({\bf{q}},\omega) U({\bf{q}},\omega)
\end{equation}
where $\delta n({\bf{q}},\omega)$ is the Fourier transform of a density modulation $\delta n({\bf{r}},t)$ induced by an external time-dependent field $U({\bf{r}},t)$, with Fourier components $U({\bf{q}},\omega)$, coupled to the particle density of the system.
The density response function is related to the dynamical structure factor through the fluctuation dissipation relation:
\begin{equation}\label{eq:fluctuationdissipation}
S({\bf{q}},\omega) \propto \Im \left( \chi_{nn}({\bf{q}},\omega + i \eta) \right), \quad \eta = 0^+
\end{equation}
For numerical and graphical purposes, a broadening factor $\eta = 10^{-3} \, t$ has been used in generating the data shown in panels b and c of Fig.~\ref{fig:Sqomega}.
Within GRPA, beyond mean-field effects are taken into account by including an interaction-driven renormalization of the external field $U({\bf{r}},t)$ due to the self-consistent dynamics of the local density and of the local pairing order parameter.
In addition, we compute $S({\bf{q}},\omega)$ within the ``text-book'' mean field BCS approximation relying on a simple uniform pairing order parameter (see e.g. \cite{PhysRevA.80.043612} ). Finally, we complement the analysis by including the dynamical structure factor for a collection of non-interacting fermions moving in an optical lattice at half-filling, which we denote $S_0({\bf{q}},\omega)$ and which is given by:
\begin{equation}
\label{fermigassqo}
\begin{split}
&
S_0({\bf{q}},\omega) = \frac{1}{2\pi^2} \int_{[-\pi,\pi]^2} d{\bf{k}} \, \delta\left( \omega - \left( \varepsilon({\bf{k}}+{\bf{q}}) - \varepsilon({\bf{k}}) \right) \right) \\
& \left\{ \theta(\varepsilon_F - \varepsilon({\bf{k}})) \left( 1 - \theta(\varepsilon_F - \varepsilon({\bf{k}} + {\bf{q}})) \right) \right\}
\end{split}
\end{equation}
where $\varepsilon({\bf{k}}) = -2 t \left( \cos k_x + \cos k_y \right)$ is the dispersion relation of the Hubbard model, while $\varepsilon_F$ is the Fermi energy, which vanishes at half-filling due to particle-hole symmetry.

The results are shown in Fig.~\ref{fig:Sqomega}, where we show the non-interacting dynamical structure factor (panel a), the BCS approximation (panel b), the GRPA result (panel c) and the QMC calculation of $S({\bf{q}},\omega)$ (panel d). 
The momentum runs along the triangle in the first Brillouin zone defined by the vertices $(0,0)$, $(0,\pi)$ and $(\pi,\pi)$. The non-interacting dynamical structure factor captures all the particle-hole excitations allowed by the dispersion relation $\varepsilon({\bf{k}})$; we stress that, at half-filling, the Fermi surface is defined by the equation $\varepsilon({\bf{k}})=0$, it has a characteristic diamond shape
and it has the well known nesting vector $(\pi,\pi)$.
This is the reason for the fact that $S_0((\pi,\pi),\omega=0)$ does not vanish: it is possible to transfer momentum $(\pi,\pi)$ at zero energy cost. This generates an instability towards a state of modulated density, which becomes stable as soon as we switch the interaction on. 

In the BCS results (panel b), we clearly see the forbidden region $\omega < 2\Delta_{BCS}$, where $\Delta_{BCS}$ is the mean-field superfluid gap of the system, that is the binding energy of the Cooper pairs. In our regime we find $\Delta_{BCS} = 1.38 \, t$, which is a parameter used in the GRPA calculations as well. For $\omega > 2 \Delta_{BCS}$, we see a quasi-particle pair continuum, closely related to the non-interacting particle-hole continuum.

In panel c we show the GRPA results, obtained using the method in \cite{PhysRevA.80.043612}.
Within this approach, an explicit time dependent perturbation coupled to both density and pairing is introduced in the mean field description, and this gives rise to a self-consistent dynamics of density and pairing order parameters which renormalizes the external fields, thus introducing in the response function effects beyond mean field. Finally, in panel d, we show the QMC results, where the momentum is discretized due to the finite size of the system.

At low energy, the superfluid gap prohibits the formation of particle-hole pairs while we observe a well defined Nambu-Goldstone collective mode $\omega({\bf{q}})$.
The agreement between QMC calculation and the GRPA calculation of the peak of the low energy mode is impressive, establishing that GRPA is a very robust method to study collective modes in superfluid fermionic systems.
For ease of comparison, the dispersion relation of the mode as predicted by GRPA is superimposed to the QMC calculation of $S({\bf{q}},\omega)$ as a filled green line.
This mode has a clear phononic shape and displays a ``roton'' minimum at ${\bf{q}}=(\pi,\pi)$, which vanishes in the thermodynamic limit and corresponds to the checkerboard crystalline structure of the system. We comment that the existence such a roton minimum had been pointed out in the early nineties relying on random phase approximation \cite{PhysRevB.46.11025}.
Our QMC data strongly suggest that $\omega({\bf{q}}) \to 0$ also as ${\bf{q}} \to 0$, although the smallest momentum that we considered is ${\bf{q}} = (\frac{\pi}{6},0)$ due to the finite size of the lattice.
While within GRPA the Nambu-Goldstone mode does not have a physical broadening and the broadening in the Figure is only due to choice of a finite $\eta$ in \eqref{eq:fluctuationdissipation} and to the discretization of frequency space, we observe that the mode $\omega({\bf{q}})$, as computed within QMC, has a broadening, with a width of the order of $0.5 \, t$, which arises from a combination of the possible decay of the mode into the quasi-particle pair continuum, in particular close to the maxima, and of artifacts due to the analytic continuation method.

\begin{figure*}[tbp]
\begin{center}
\includegraphics[width=\textwidth]{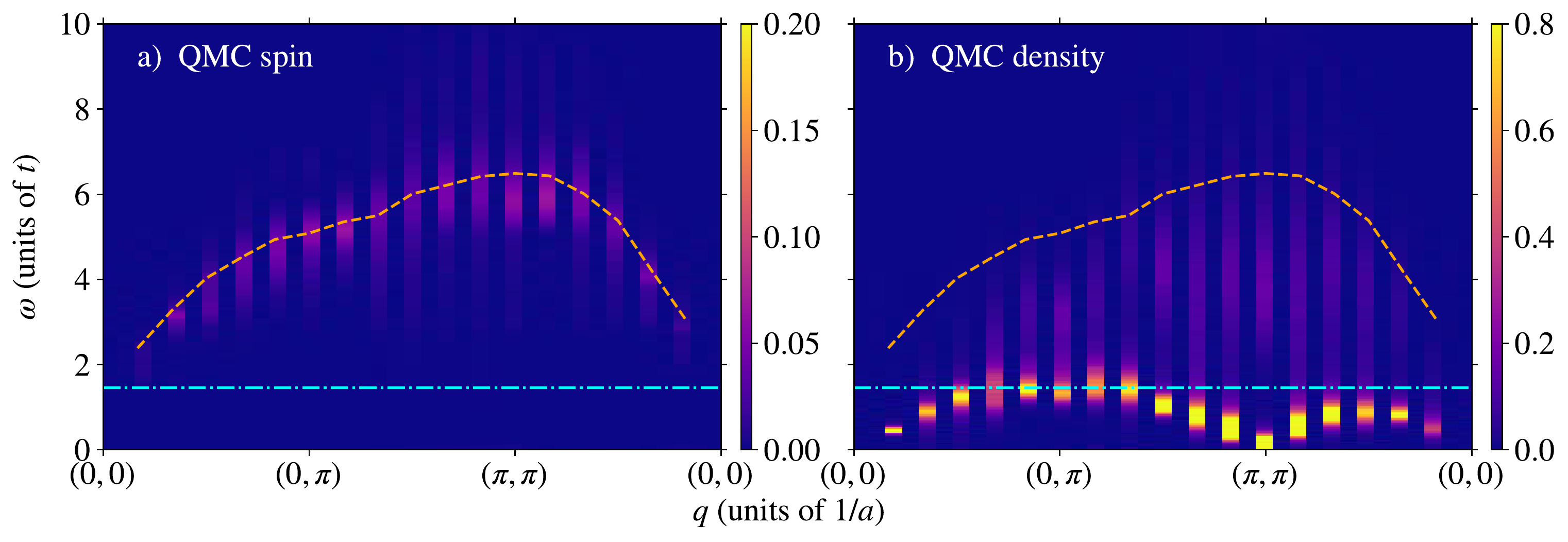}
\caption{(Color online) Panel (a): color plot of the dynamical spin structure factor $S^s({\bf{q}}, \omega)$ (arbitrary units) for the attractive Hubbard model with $U/t=-4$ at half-filling as a function of momentum ${\bf{q}}$ along a triangle in the first Brillouin zone and frequency $\omega$, in units of the hopping amplitude $t$. Panel (b): color plot of the dynamical structure factor $S({\bf{q}}, \omega)$ (arbitrary units) for the same system calculated with QMC. In both panels the dot-dashed horizontal line is twice the quasi-particle gap calculated with QMC from Ref.~\cite{PhysRevB.94.085140}, while the dashed line is the center of the spin mode $\omega_s({\bf{q}})$, defined in \eqref{omegas}.}
\label{fig:Sqomegaspin}
\end{center}
\end{figure*}

At higher energy, the QMC calculation shows a significant difference with respect to the GRPA result. The latter, above the energy $2\Delta_{BCS}$
clearly displays a quasi-particle pair continuum, with a shape which is qualitatively very similar to the non-interacting particle-hole continuum and to the simple BCS result.
The QMC result for $S({\bf{q}},\omega)$ at higher energy appears to be more complex. While the resolution here is certainly limited due to the difficulty of performing the analytic continuation and to the finite size of the system, we can still extract some important physical observations.
The first observation is that the spectral weight in the higher energy branch in the QMC result exists at energies significantly lower than $2\Delta_{BCS}$: this can be interpreted as a renormalization of the superfluid gap due to the many-body correlations beyond GRPA $\Delta < \Delta_{BCS}$, which is consistent with the calculations of the charge gap in repulsive models \cite{PhysRevB.94.085140}.
The value for $\Delta$ that can be obtained with a QMC technique taking the size of the system and the boundary conditions into account, is shown as a horizontal dotted-dashed cyan line superimposed to panel d. The value we use is $\Delta=0.73 \, t$ \cite{PhysRevB.94.085140}.
Also, even the maximum of $S({\bf{q}},\omega)$ appears to be moved to significantly lower energy.
Perhaps the most natural way to interpret the QMC data is to view the high-energy branch as quasi-particle pair continuum which is strongly renormalized by many-body correlations beyond GRPA. 
We observe also that the QMC higher energy branch appears to be less broad with respect to the RPA calculation, which may suggest some form of coherent behavior. We do not have the resolution here to make a strong statement about a possible second collective mode, although the possibility of its existence is intriguing and will stimulate us to look for some more sensitive probes to investigate further.
We find it important to mention that the existence of two modes can be expected in a supersolid \cite{PhysRevLett.108.175301,saccani_bose_2011} if we consider that the system is breaking at the same time discrete translational symmetry of the lattice, giving rise to the crystalline order, and $U(1)$ symmetry, giving rise to the pairing order.
Also, it is impossible not to think about the possibility to detect a Higgs mode related to breaking of $U(1)$ and corresponding to fluctuations of the amplitude of the complex pairing order parameter. It is generally expected that such a mode would quickly decay into the quasi-particle pair continuum and thus it cannot be detected as a well defined peak in the dynamical structure factor in this regime. Anyway, as far as we know, unbiased calculations of $S({\bf{q}},\omega)$ are just starting to appear and this makes it very interesting to actually look for the Higgs mode. It will be very exiting, in future work, to explore different dynamical correlations which may filter out the signal more efficiently than the dynamical structure factor.

We complement our investigation of the dynamical properties of the system with a calculation of the spin dynamical structure factor:
\begin{equation}
\label{SecII-Sqos}
S^s({\bf{q}},\omega) = \frac{1}{ N} \int_{-\infty}^{+\infty} \frac{dt}{2\pi} \, e^{i\omega t}
 \left\langle \Psi_0 \, | e^{i t \hat{H}} \hat{n}^s_{{\bf{q}}} e^{- i t \hat{H}}\, \hat{n}^s_{{-\bf{q}}} | \Psi_0 \right\rangle 
\end{equation}
where the Fourier component of the spin density is given by:
\begin{equation}
\hat{n}^s_{{\bf{q}}} = \frac{1}{2} \sum_{{\bf{k}}} \left( \hat{c}^{\dagger}_{{\bf{k}},\uparrow} \hat{c}^{}_{{\bf{k}}+{\bf{q}},\uparrow} -\hat{c}^{\dagger}_{{\bf{k}},\downarrow} \hat{c}^{}_{{\bf{k}}+{\bf{q}},\downarrow} \right) \,.
\end{equation}
This correlation function gives us access to the spectrum of spin density fluctuations of the attractive Hubbard model at half-filling. Our QMC result is shown in panel a of Fig.~\ref{fig:Sqomegaspin}, where the spin dynamical structure factor is compared with the density dynamical structure factor. 
The first observation is that $S^s({\bf{q}},\omega)$ is zero for $\omega < 2\Delta$. This can be understood by observing that the creation of a spin density modulation requires breaking pairs, as the pairing mechanism ``locks'' the involved particles in singlet on-site pairs, thus preventing the formation of spin density waves. Incidentally, this is the reason why we do not show a comparison with GRPA for $S^s({\bf{q}},\omega)$: GRPA strongly relies on BCS superfluid gap which is strongly overestimated. We thus expect that the spin dynamical correlations within GRPA will be consistently shifted to higher energies.
When $\omega > 2\Delta$, in $S^s({\bf{q}},\omega)$ we observe that the spectral weight is concentrated around a somehow broad but well defined mode which, very interestingly, appears to be consistent with the upper limit of the quasi-particle pair continuum in $S({\bf{q}},\omega)$ (panel b in Fig.~\ref{fig:Sqomegaspin}). The center of mass of the spectral weight in the spectrum of spin fluctuations is given by:
\begin{equation}
\label{omegas}
\omega_s({\bf{q}}) = \frac{m_1^s({\bf{q}})}{m_0^s({\bf{q}})} = \frac{\int_{0}^{+\infty} d\omega \, \omega \, S^s({\bf{q}},\omega)}{\int_{0}^{+\infty} d\omega \, S^s({\bf{q}},\omega) }
\end{equation}
where the denominator is just the Fourier transform of the local spin density correlation:
\begin{equation}
m_0^s({\bf{q}}) = \frac{1}{N} \left\langle \Psi_0 \, | \hat{n}^s_{{\bf{q}}} \, \hat{n}^s_{{-\bf{q}}} | \Psi_0 \right\rangle 
\end{equation}
which is an output of the QMC calculation, while the numerator can be computed using the (spin) $f$-sum rule. The calculation is almost identical to the one for the density, which we detail in Appendix A, including also the result for the spin dynamics. 
In Fig.~\ref{fig:Sqomegaspin} we have superimposed $\omega_s({\bf{q}})$ as a dashed line to both the spin and the density dynamical structure factors.
It is evident that $\omega_s({\bf{q}})$ lies close to the upper limit of the quasi-particle pair continuum uniformly as a function of ${\bf{q}}$, which explains the broadness of the spin mode, that can easily decay into excitations in the continuum. Interestingly, for momenta close to ${\bf{q}}=(0,0)$ (considering limitations due to the finite size of the system), the spin mode appears to be more narrow and to be present in a region where, in the density dynamical correlation, the spectral weight of the quasi-particle pair continuum is very low. We find it relevant to mention that, in a recent interesting paper \cite{Zhao_2020}, Zhao et al., studying a dilute Fermi gas, interpret a similar spin mode as the celebrated Higgs mode. In our opinion, the precise connection with the Higgs mode is hard to establish relying on our results here, but the possibility is very intriguing and we are poised to investigate further in future works. A precise study of the low momentum behavior of $\omega_s({\bf{q}})$ will be needed, which requires larger lattices, and a theoretical study aimed at establishing which dynamical correlation is expected to be able to detect the Higgs mode: the latter is expected to appear as a peak around $2\Delta$, but it may easily decay into quasi-particle pairs which populate these energies.

Before drawing our conclusions, we mention that the study of dynamical two-body susceptibilities is extremely relevant also in the realm of the repulsive two-dimensional Hubbard model, in connection with the physics of Cuprates. The reader can have an overview by looking at the recent papers \cite{PhysRevB.100.075123,PhysRevB.100.235107} and the several references therein, summarizing a vast body of literature. For experiments on Cuprates, and in particular on parent compounds which are commonly modeled using the Hubbard hamiltonian at half-filling, the reader can look for example at the paper \cite{Coldea_SpinWavesElectronic_2001} and references therein. In that context, the majority of studies focus on the magnetic dynamical susceptibility \cite{Peres_Spinwavedispersion2CuO_2002,Katanin_SpinexcitationsLa2CuO4_2002,PhysRevB.100.075123}, which, at half-filling, can be mapped into the dynamical structure factor of the attractive model which is the focus of our work. Most approaches focus on main coherent excitation, while the higher energy excitations are not usually addressed, so we expect that our data can provide important information. On the other hand, we are aware of only a few works \cite{PhysRevB.100.235107} addressing the study of the dynamical charge susceptibility which can be mapped to the spin dynamical structure factor of the attractive model, which we computed in this paper. Random phase approximation has been extensively used for repulsive models, and other very promising more recent techniques have been used as well like determinantal QMC, dynamical mean field theory and its generalizations, like dynamical cluster approximation and the technique of dual fermions. Determinantal QMC is unbiased but it is limited to high temperatures \cite{PhysRevB.100.075123}, and so our results can provide a new important benchmark also for repulsive models, as we work at $T=0$ K. On the other hand, dynamical cluster approximation is an approximation to the self-energy introducing coarse graining in momentum space, which limits momentum resolution, while the dual fermion technique treats the self-energy with an innovative resummation scheme, which recovers continuum momentum dependence: it is expected to be accurate at high temperature but it is known to be uncontrolled (see \cite{PhysRevB.100.075123} for the details about these techniques). 
All these advanced computational methodologies work at finite temperature, the low temperature regime being very challenging and thus we expect that our results at zero temperature may provide useful benchmarks for them.

\section{Conclusions}
We used an exact methodology to sample the ground state and the imaginary time propagator of the attractive Hubbard model at half-filling. Our central result was the study of the spectrum of density correlations of the system. We compared our unbiased predictions with the Generalized Random Phase approximation, which is widely considered a method of choice for Fermi superfluids. The accuracy of the latter, in fact, is very hard to assess in theoretical approaches due to the difficulty of considering the terms beyond GRPA, and the availability of unbiased numerical results can provide a very valuable benchmark.
We were able to compute exactly the intermediate scattering function in imaginary time and we used the state-of-art GIFT method to perform the analytic continuation necessary to obtain an estimation of the dynamical structure factor $S({\bf{q}},\omega)$.
The physical regime is very interesting from the point of view of the study of dynamical density fluctuations, due to the complex broken symmetries. We observe a fermionic supersolid, with long-range order both in the density correlations, corresponding to a crystalline checkerboard order with wave vector ${\bf{q}}=(\pi,\pi)$ and in the on-site $s$-wave pairing correlations, corresponding to a superfluid phase.  
Our QMC calculations agree very well with the GRPA results in detecting a narrow low-energy Nambu-Goldstone mode $\omega({\bf{q}})$ which clearly vanishes at ${\bf{q}}=(\pi,\pi)$ as a consequence of the checkerboard density modulation.
The higher energy branch shows discrepancies between the QMC and the GRPA calculation, with the QMC upper branch appearing at significantly lower energy and being less diffuse. The simplest interpretation is a renormalization of the superfluid gap and of the quasi-particle pair continuum due to the many-body correlations beyond GRPA. Nevertheless, the more intriguing possibility of a signature of a more coherent mode, maybe related to the Higgs mode, cannot be excluded at this time and we are poised to explore different dynamical correlations that may help us filter the signal more efficiently. We studied also the spectrum of the spin density fluctuations of the system, and we found a well defined spin mode with energy close to the upper limit of the quasi-particle pair continuum. The low momentum behavior of this mode is intriguing and may be another indication of the possibility to detect the Higgs mode in this system, although this certainly requires further extensive investigation.

In summary, our results provide a robust confirmation of the reliability of GRPA as a method of choice for the calculation of low-energy collective modes in fermionic systems. GRPA is certainly more efficient than QMC in studying large systems due to the linear scaling in the system size, as opposed to the still favorable cubic scaling of the QMC technique.
Moreover, GRPA has the tremendous advantage of working directly in real time, thus not requiring the delicate analytic continuation. Nevertheless, QMC yields unbiased results, at least in imaginary time, and we have now evidence that some important effects beyond GRPA may provide significant new insight in the physics of the systems. The results encourage us to further investigate the possibility to interface GRPA and QMC, potentially through the introduction of effective parameters, following an idea similar to the one presented in \cite{Vitali2019}.

Finally, we plan to investigate, in future studies, the behavior of the model as we move away from half-filling. In particular, we would like to allow the particle density to change, in order to see if there may be a density modulated phase which becomes stable at some values of the filling, and to break the spin balance, which is expected to allow the system to develop highly nontrivial pairing mechanisms, related to exotic phases like the Fulde-Ferrel-Larkin-Ovchinnikov phase.

One of us, E.V., acknowledges useful discussions with Shiwei Zhang, Peter Rosenberg, Hao Shi and Yuan-Yao He. Computing was carried out at the Extreme Science and Engineering Discovery Environment (XSEDE), which is supported by National Science Foundation grant number ACI-1053575. D.E.G. acknowledges the CINECA awards IscraB PANDA and IscraC RENNA for the availability of high performance computing resources and support.

\appendix

\section{The \texorpdfstring{$f$}{f}-sum Rule for the Hubbard Model}
\label{appA}

	In this appendix we will present some details about the analytic calculation of the $f$-sum rule for the Hubbard Hamiltonian. The main purpose is to obtain a simple expression for the first moment:
\begin{equation}
m_1({\bf{q}}) = \int_{0}^{+\infty} d\omega \, \omega S({\bf{q}},\omega)
\end{equation}
the following exact expression holds:
\begin{equation}
\label{m1q}
m_1({\bf{q}}) = \frac{1}{2N} \left\langle \, \sum_{{\bf{k}},\sigma} \left( \varepsilon({\bf{k}}+{\bf{q}}) + \varepsilon({\bf{k}}-{\bf{q}}) \right) 
 \hat{c}^{\dagger}_{{\bf{k}},\sigma} \hat{c}^{}_{{\bf{k}},\sigma} - 2 \hat{T}\, 
\, \right\rangle
\end{equation}
where:
\begin{equation}
 \hat{T} = \sum_{{\bf{k}},\sigma} \varepsilon({\bf{k}}) \hat{c}^{\dagger}_{{\bf{k}},\sigma} \hat{c}^{}_{{\bf{k}},\sigma} 
\end{equation}
is the kinetic energy operator of the Hubbard model, with the usual dispersion relation:
\begin{equation}
 \varepsilon({\bf{k}}) = -2 t \sum_{i=1}^{d} \cos(k_i)
\end{equation}
with $d$ being the dimensionality.

\begin{figure}[tbp]
\begin{center}
\includegraphics[width=\columnwidth, angle=0]{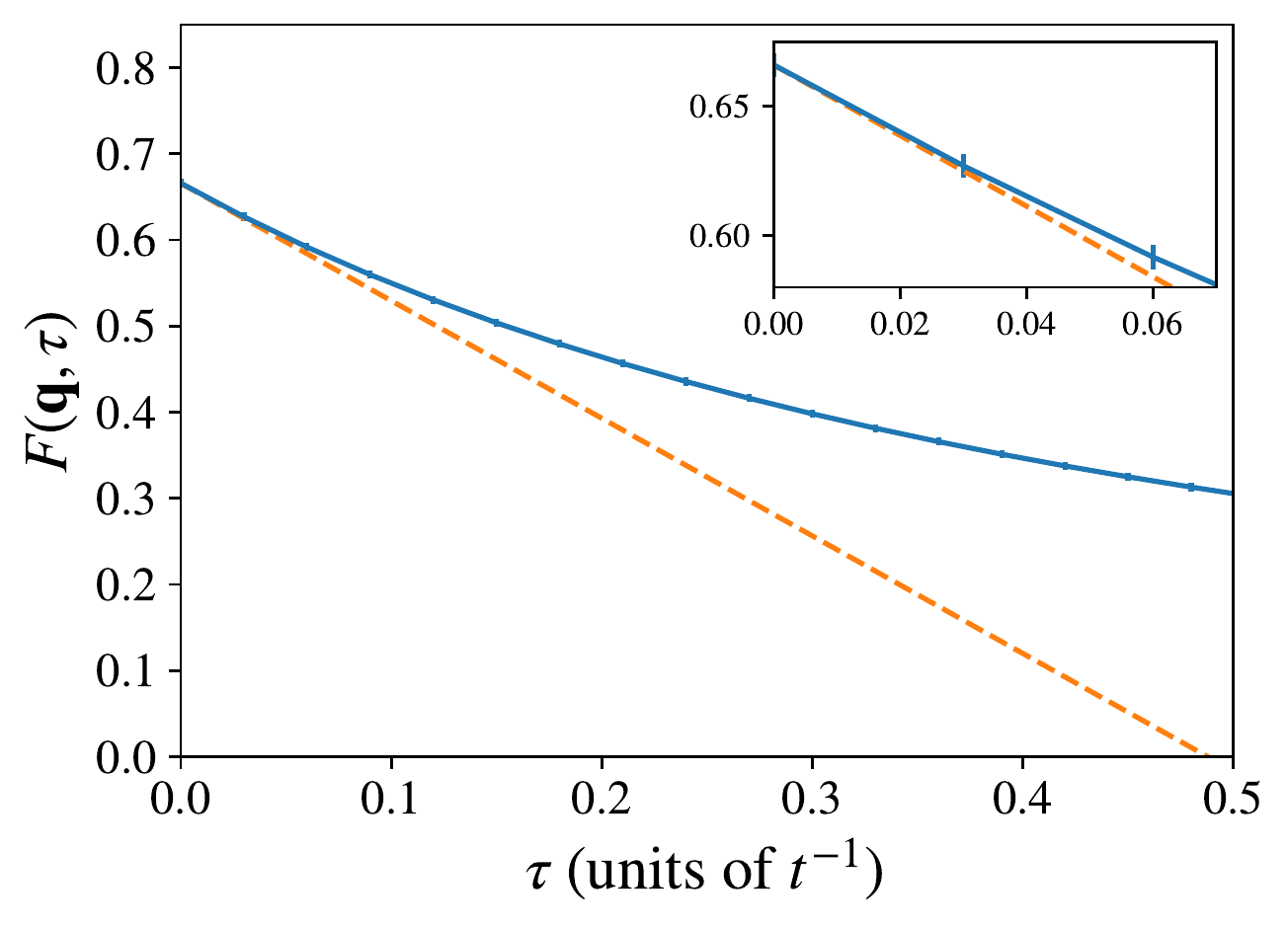}
\caption{(Color online) For ${\bf{q}}=\frac{2\pi}{L}(3,3)$ we show the comparison between the intermediate scattering function (dimensionless) as computed from QMC (blue dots with errorbars connected by full line to guide the eye) and the straight line $y(\tau) = F({\bf{q}},\tau=0) - \tau m_1({\bf{q}})$. We see that the line $y(\tau)$ is clearly tangent to $F({\bf{q}},\tau)$ at $\tau=0$. The inset is just a zoom in.}
\label{fig:slope}
\end{center}
\end{figure}

We stress that the result \eqref{m1q} critically depends on the form of the dispersion relation and it reduces to the traditional $f$-sum rule when the quadratic dispersion is used. We also comment that \eqref{m1q} cannot be evaluated analytically as this would require the calculation of the spin resolved momentum distribution: $n({\bf{k}},\sigma) = \langle \,\hat{c}^{\dagger}_{{\bf{k}},\sigma} \hat{c}^{}_{{\bf{k}},\sigma} 
\, \rangle$; however this quantity can be easily obtained from a QMC simulation.

In order to derive \eqref{m1q} we start from the expression:
\begin{equation}
\label{sqo}
S({\bf{q}},\omega) = \frac{1}{ N} \sum_n \delta\left( \omega - (E_n - E_0 ) \right)
\left| \left\langle \Psi_0 \, | \hat{n}_{{\bf{q}}} \, | \Psi_n \right\rangle \right|^2
\end{equation}
where $\{| \Psi_n \rangle\}$ is an orthonormal basis of eigenvectors of the Hamiltonian related to the eigenvalues $\{E_n\}$. Also:
\begin{equation}
\hat{n}_{{\bf{q}}} = \sum_{{\bf{k}}, \sigma} \hat{c}^{\dagger}_{{\bf{k}},\sigma} \hat{c}^{}_{{\bf{k}}+{\bf{q}},\sigma}
\end{equation}
is the Fourier transform of the local density of the particles.

From \eqref{sqo} we immediately conclude that:

\begin{equation}
m_1({\bf{q}}) = \frac{1}{ N} \sum_n (E_n - E_0 ) \left| \left\langle \Psi_0 \, | \hat{n}_{{\bf{q}}} \, | \Psi_n \right\rangle \right|^2 \,.
\end{equation}
This equality through simply algebraic manipulations can be recast as:
\begin{equation}
m_1({\bf{q}}) = \frac{1}{2N} \left\langle \, \left[ \left[ \hat{n}_{{\bf{q}}} \,, \, \hat{H} \right] \,, \, \hat{n}_{{-\bf{q}}} \right] 
\, \right\rangle \,.
\end{equation}
The main result \eqref{m1q} follows from the simple but lengthy explicit evaluation of the commutators.

The possibility to compute the first moment $m_1({\bf{q}}) $ is important for the analytic continuation procedure, as it allows us to improve the constraints in the GIFT method.
The relation between the $f$-sum rule and the intermediate scattering function rests on the following equality:
\begin{equation}
\label{slope}
\frac{\partial}{\partial \tau} F({\bf{q}},\tau=0) = - m_1({\bf{q}}) \,.
\end{equation}
We find it useful to show in Fig.~\ref{fig:slope} that this relation is satisfied by our QMC calculations, this being a further cross-check for the accuracy of QMC. In fact, although there is no sign problem, making the results unbiased, the calculation of $F({\bf{q}},\tau) $ requires the sampling of the imaginary time evolution, while $ m_1({\bf{q}})$ just requires static calculations, which are more straightforward. Relation \eqref{slope} provides a useful additional verification of the accuracy of the evaluation of $F({\bf{q}},\tau)$.
An identical calculation for the spin dynamical structure factor yields:
\begin{equation}
m_1^s({\bf{q}}) = \frac{m_1({\bf{q}}) }{4} \quad.
\end{equation}

%


\end{document}